\documentclass{webofc}
\usepackage[varg]{txfonts}   
\usepackage{amsmath}
\usepackage{float}
\usepackage{subfigure}
\usepackage{bm}

\begin{document}

\title{Collective flow of light nuclei and hyper-nuclei in Au+Au collisions at $\sqrt{s_{NN}}$ = 3, 14.6, 19.6, 27, and 54.4 GeV using the STAR detector}

\author{\firstname{Rishabh} \lastname{Sharma}\inst{1}\fnsep\thanks{\email{rishabhsharma@students.iisertirupati.ac.in}},
        \firstname{} \lastname{for the STAR Collaboration}
}

\institute{Indian Institute of Science Education and Research (IISER) Tirupati}

\abstract{%
  Light nuclei and hyper-nuclei are produced in abundance in heavy-ion collisions. The production mechanism of these species in heavy-ion collisions still remains to be understood. In these proceedings, we report the transverse momentum and centrality dependence of elliptic flow ($v_2$) of $d$, $t$, and $^3\text{He}$ in Au+Au collisions at $\sqrt{s_{NN}}$ = 14.6, 19.6, 27, and 54.4 GeV. Mass number scaling of $v_2(p_T)$ of light nuclei is shown. We also report the first observation of hyper-nuclei $^{3}_{\Lambda}$H and $^{4}_{\Lambda}$H directed flow ($v_1$) in $\sqrt{s_{NN}}$ = 3 GeV mid-central (5-40\%) Au+Au collisions in the fixed target mode. 
}
\maketitle
\section{Introduction}
\label{intro}
The production and interaction of light nuclei and hyper-nuclei in high-energy heavy-ion collisions have been a focus of theoretical and experimental interests for a long time. The production of light nuclei in heavy-ion collisions can be explained by the coalescence of produced or transported nucleons \cite{RefCoal1,RefCoal2,RefCoal3,RefCoal4,RefCoal5}. Due to their low binding energies, it is more likely that they are formed at later stages of the evolution of the fireball. Therefore, studying the collective flow of light nuclei and hyper-nuclei can provide insights into their production mechanism in heavy-ion collisions. Further, the study of the collective flow of hyper-nuclei will shed light on the hyperon-nucleon (YN) interaction in dense nuclear medium \cite{NS1,NS2}.
\\In the following sections, elliptic flow ($v_2$) of $d$, $t$, and $^3\text{He}$ in Au+Au collisions at $\sqrt{s_{NN}}$ = 14.6, 19.6, 27, and 54.4 GeV and the directed flow ($v_1$) of $\Lambda$, $^{3}_{\Lambda}$H, and $^{4}_{\Lambda}$H in mid-central Au+Au collisions at $\sqrt{s_{NN}}$ = 3 GeV are discussed.

\section{Analysis details}  
\label{sec-1}
The data presented here was collected in the phase 2 of the Beam Energy Scan (BES-II) program by the STAR experiment at RHIC. Particle identification is performed using the Time Projection Chamber (TPC) and Time of Flight (TOF) detectors. The TPC is the primary tracking detector in the STAR experiment which uses specific ionisation energy loss ($dE/dx$) in a large gas volume to detect trajectories of charged particles. TOF detector allows the identification of particles of interest using a restriction on the mass-square ($m^2$) of the particle. It helps improve particle identification in the intermediate transverse momentum region. The hyper-nuclei $^{3}_{\Lambda}$H and $^{4}_{\Lambda}$H are reconstructed from the following decay channels: $^{3}_{\Lambda}$H $\to$ $^3\text{He}$ + $\pi^-$, $^{3}_{\Lambda}$H $\to$ $d$ + $p$ +$\pi^-$, and $^{4}_{\Lambda}$H $\to$ $^4\text{He}$ + $\pi^-$. In this study, $v_2$ measurements of light nuclei are performed in Au+Au collisions in the collider mode whereas $v_1$ measurements of hyper-nuclei are performed in the fixed-target (FXT) mode. In the fixed target mode, the gold beam of 3.85 GeV/u is bombarded on a thin gold target with 1\% interaction probability. The target is located at 200 cm along the beam direction from the center of the TPC. In Au+Au collisions at $\sqrt{s_{NN}}$ = 14.6, 19.6, 27, and 54.4 GeV, the second order event plane angle is constructed using the TPC to measure the $v_2$ of $d$, $t$, and $^3\text{He}$. The $v_1$ of hyper-nuclei in Au+Au collisions at $\sqrt{s_{NN}}$ = 3 GeV is measured using the first order event plane determined by the Event Plane Detector (EPD) [8]. EPD is designed to measure the azimuthal patterns of charged particles emitted in backward and forward direction relative to the beam direction. 

\section{Results}
\label{sec-2}
\subsection{Elliptic flow of light nuclei}
Figure 1 shows $v_2$ as a function of $p_T$ in minimum bias Au+Au collisions at $\sqrt{s_{NN}}$ = 14.6, 19.6, 27, and 54.4 GeV. A monotonous increase in $v_2$ of light nuclei with $p_T$ across all four center-of-mass energies is observed. 
\begin{figure}[H]
    \centering
    \subfigure{\includegraphics[width=0.24\textwidth]{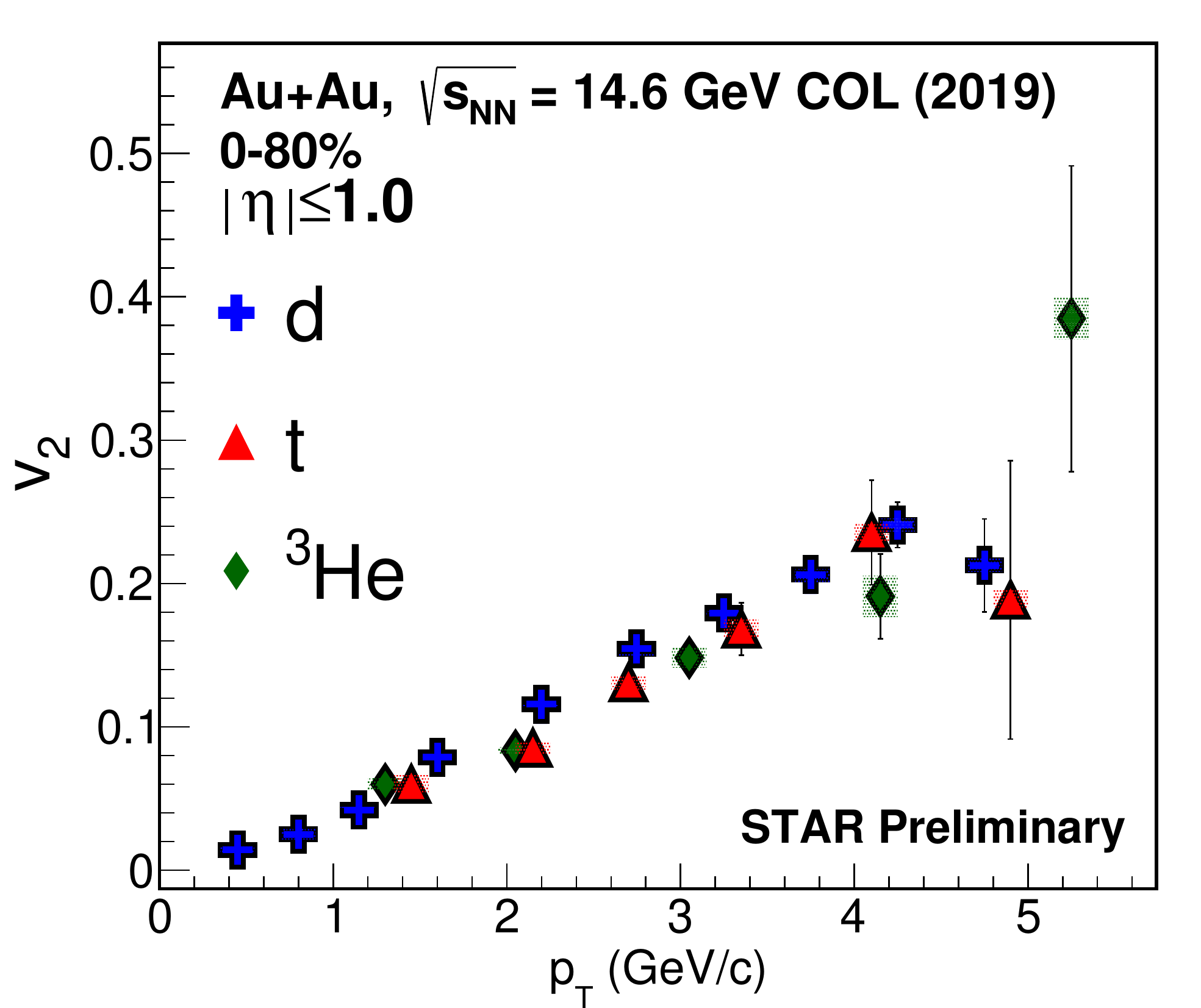}}
    \subfigure{\includegraphics[width=0.24\textwidth]{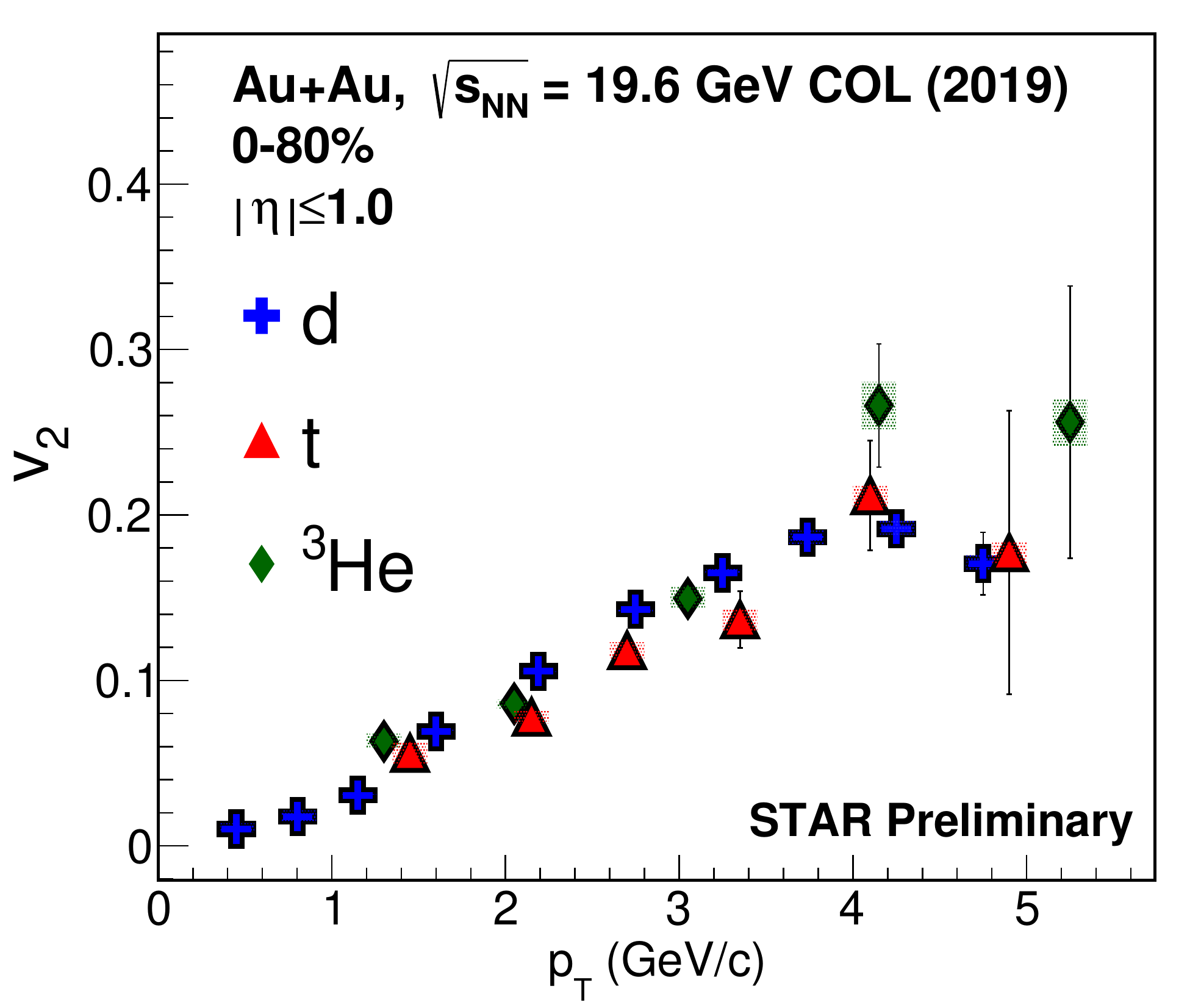}}
     \subfigure{\includegraphics[width=0.24\textwidth]{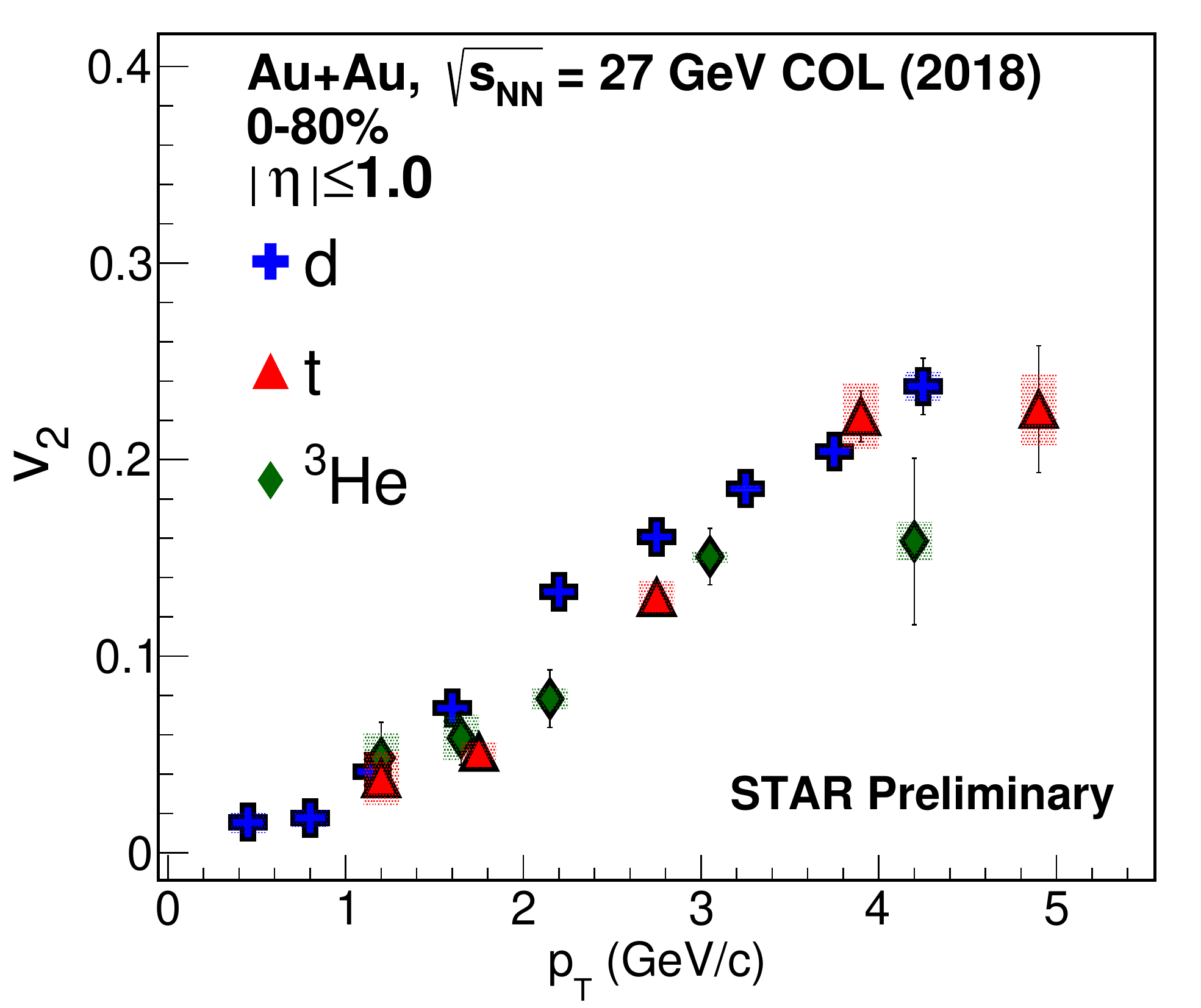}}
    \subfigure{\includegraphics[width=0.24\textwidth]{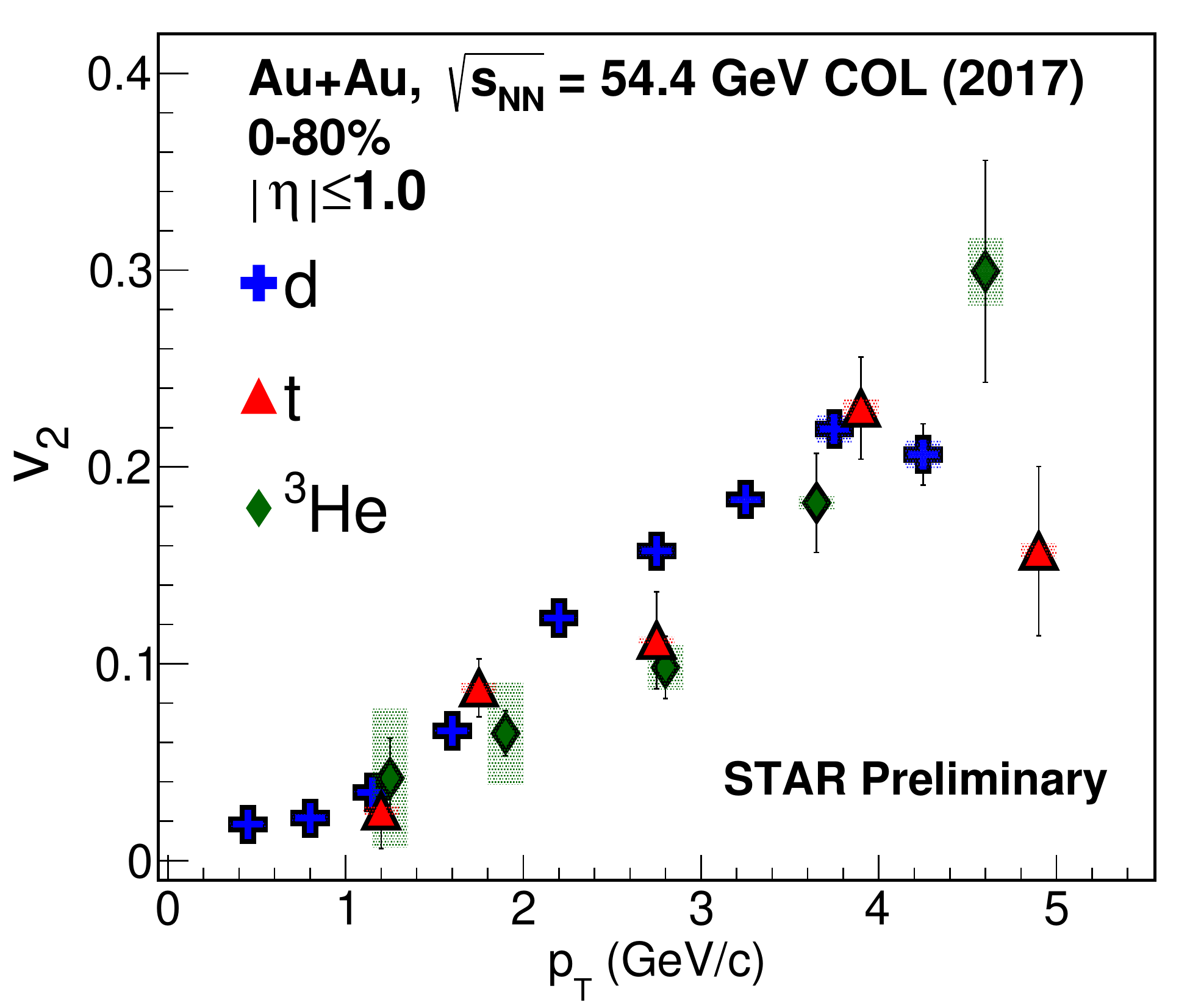}} 
    \caption{$v_2(p_T)$ of light nuclei ($d$, $t$, and $^3\text{He}$) in minimum bias Au+Au collisions at $\sqrt{s_{NN}}$ = 14.6, 19.6, 27, and 54.4 GeV. Vertical lines and shaded area at each marker represent statistical and systematic uncertainties, respectively.}
    \label{fig:foobar}
\end{figure}

Centrality dependence of $v_2$ of $d$ is shown in Fig. 2. The nuclei $v_2$ is measured in two centrality ranges 0-30\% and 30-80\% for Au+Au collisions at $\sqrt{s_{NN}}$ = 19.6, 27, and 54.4 GeV. $v_2(p_T)$ of $d$ shows a strong centrality dependence for all the three energies. It is observed that peripheral collisions have a higher $v_2$ compared to more central collisions. This can be explained by the fact that more peripheral collisions have a greater spatial anisotropy compared to the central collisions. 
\begin{figure}[H]
    \centering
    \subfigure{\includegraphics[width=0.30\textwidth]{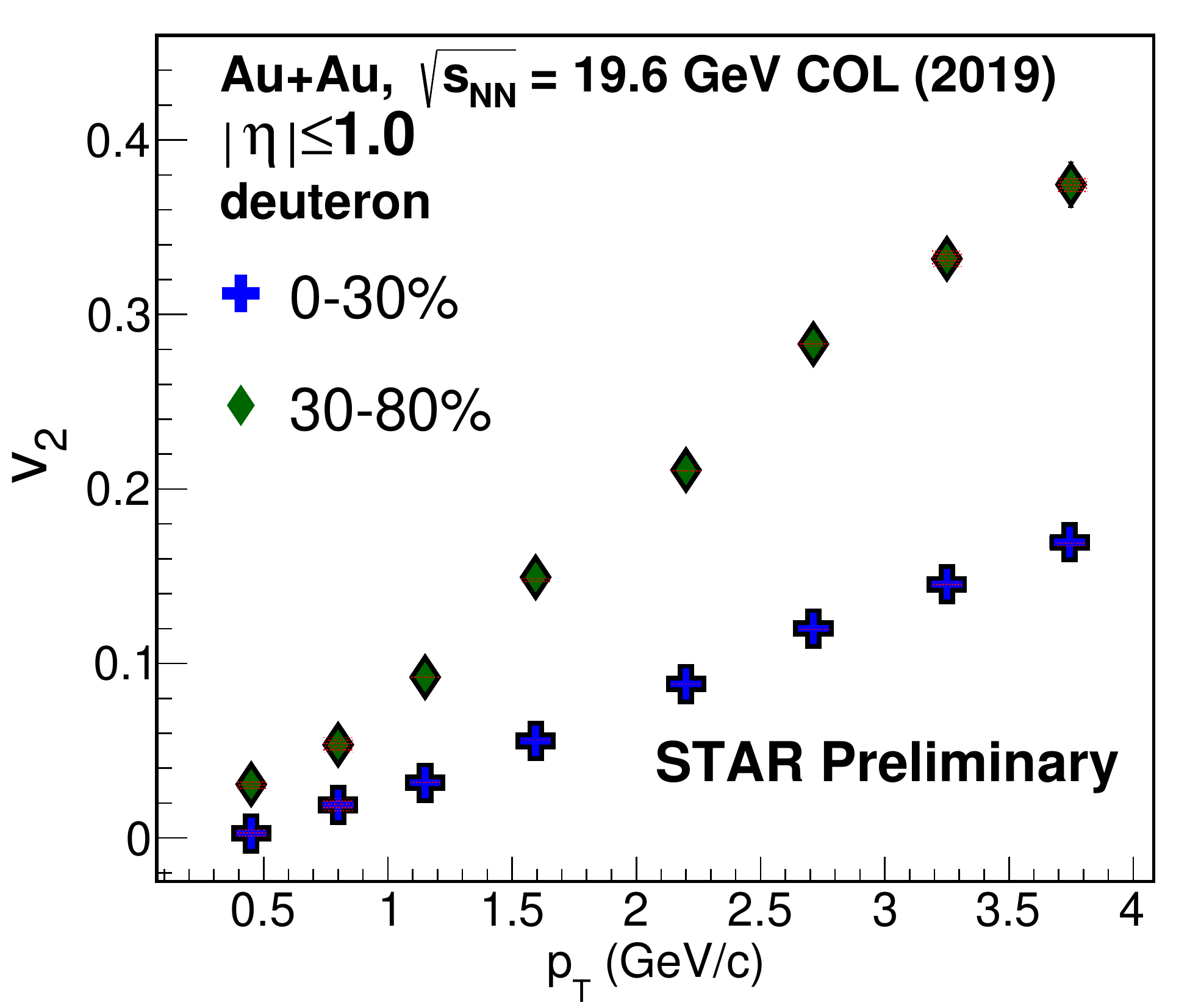}}
    \subfigure{\includegraphics[width=0.30\textwidth]{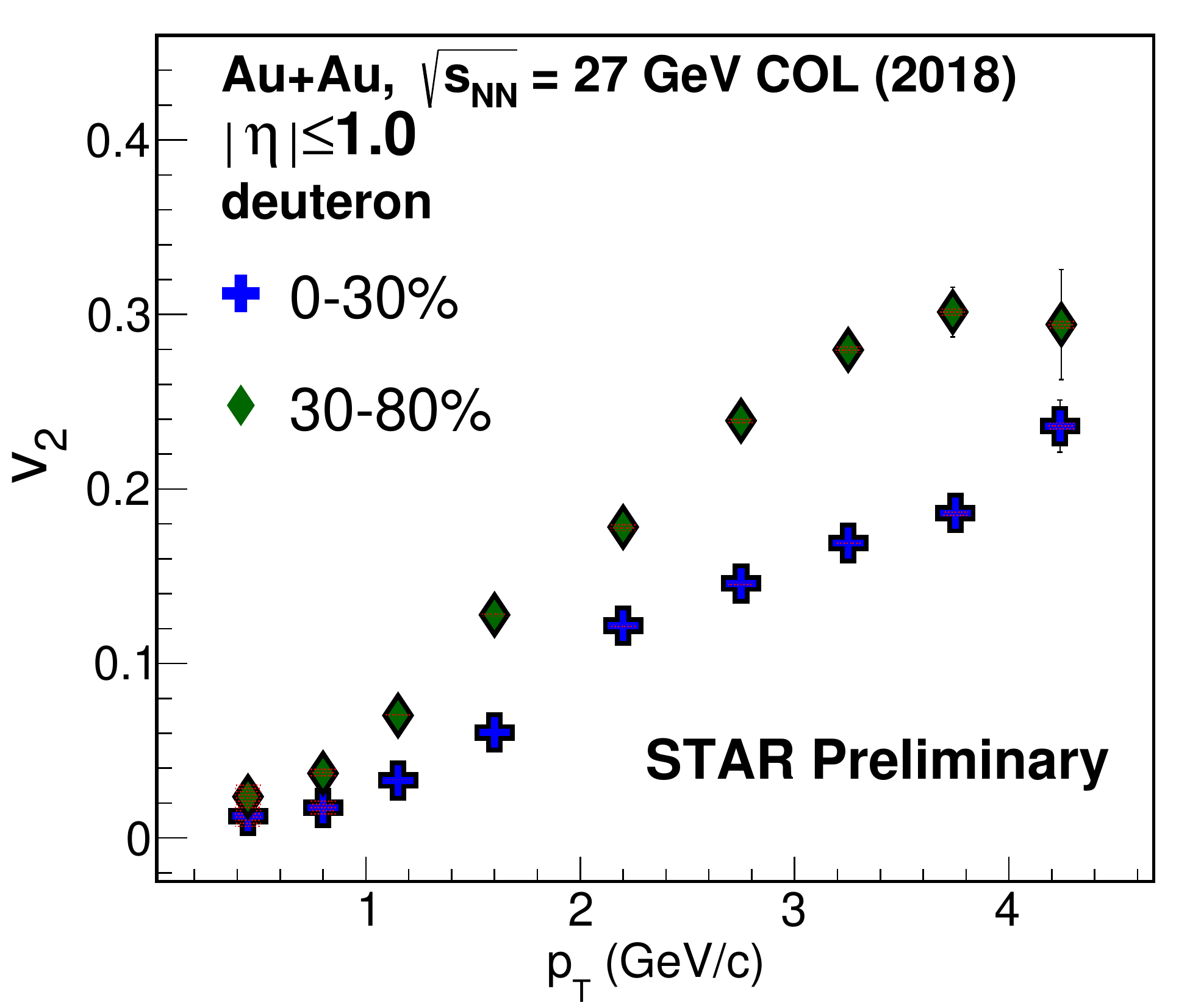}}
    \subfigure{\includegraphics[width=0.30\textwidth]{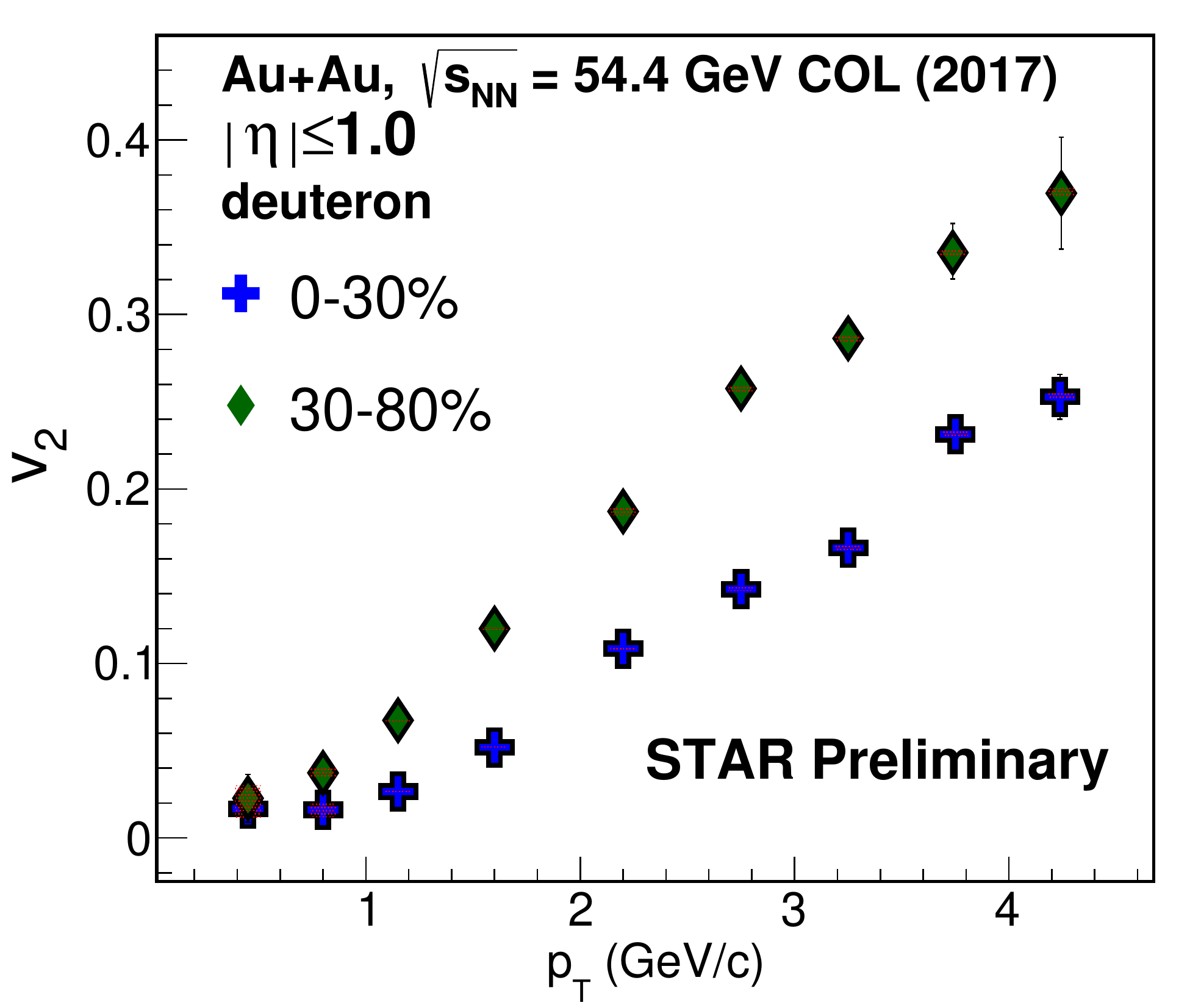}} 
    \caption{Centrality dependence of $v_2$ of $d$ as a function of $p_T$ in Au+Au collisions at $\sqrt{s_{NN}}$ = 19.6, 27, and 54.4 GeV. Vertical lines and shaded area at each marker represent statistical and systematic uncertainties, respectively.}
    \label{fig:foobar}
\end{figure}

Coalescence model predicts that if a composite nuclei is produced by coalescence of $n$ number of nucleons, very close to each other in phase-space, then the $v_2$ of the composite nuclei will be $n$-times the $v_2$ of the individual nucleons \cite{MSN1,MSN2}. This is called as mass number scaling. Figure 3 shows the comparison of light nuclei $v_2/A$ as a function of $p_T/A$ (where $A$ is the mass number of the nuclei) with proton $v_2/A$ (where $A=1$). Light nuclei $v_2$ has been observed to follow the mass number scaling within 20-30\%.
\begin{figure}
    \centering
    \subfigure{\includegraphics[width=0.24\textwidth]{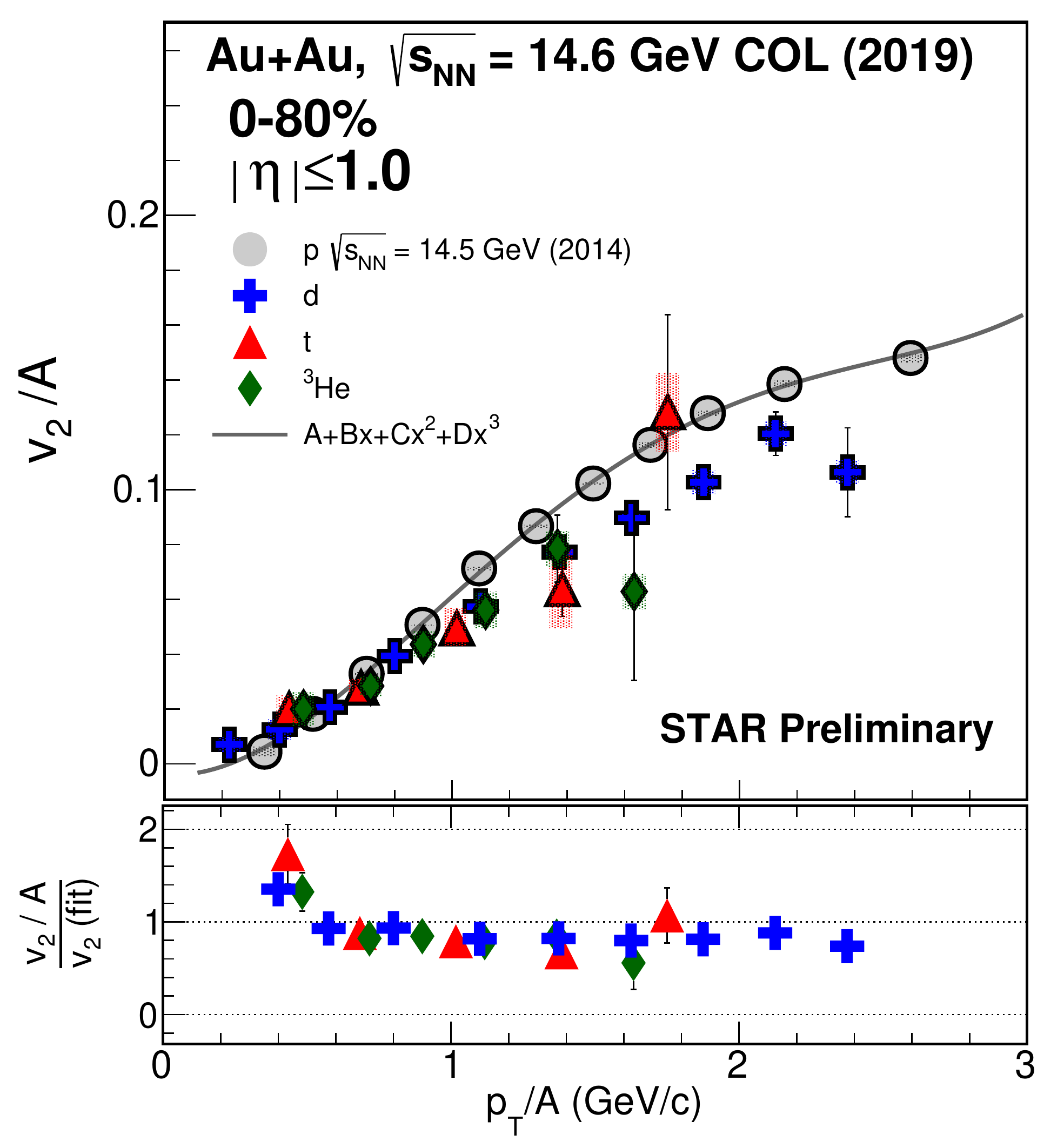}}
    \subfigure{\includegraphics[width=0.24\textwidth]{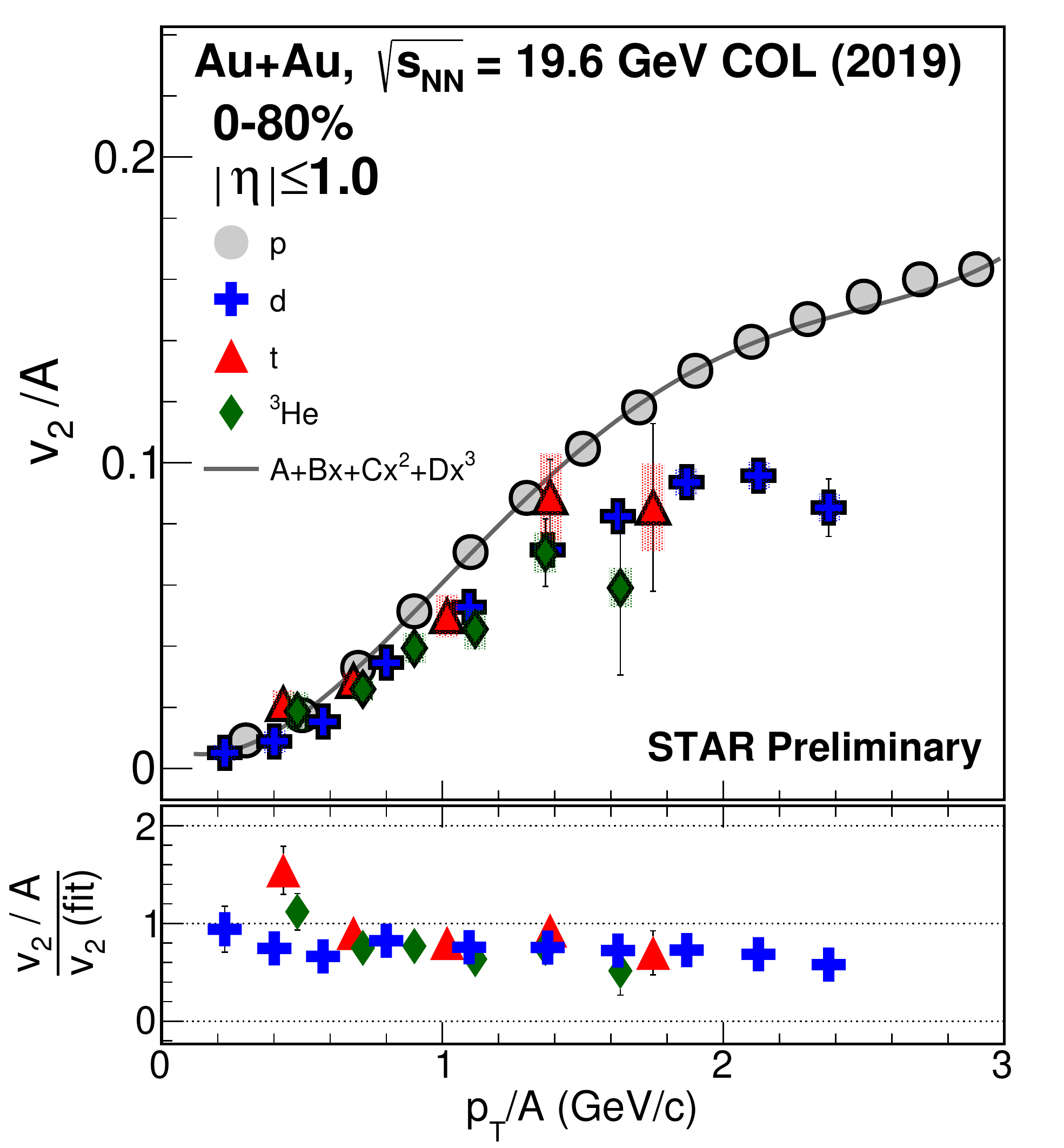}}
    \subfigure{\includegraphics[width=0.24\textwidth]{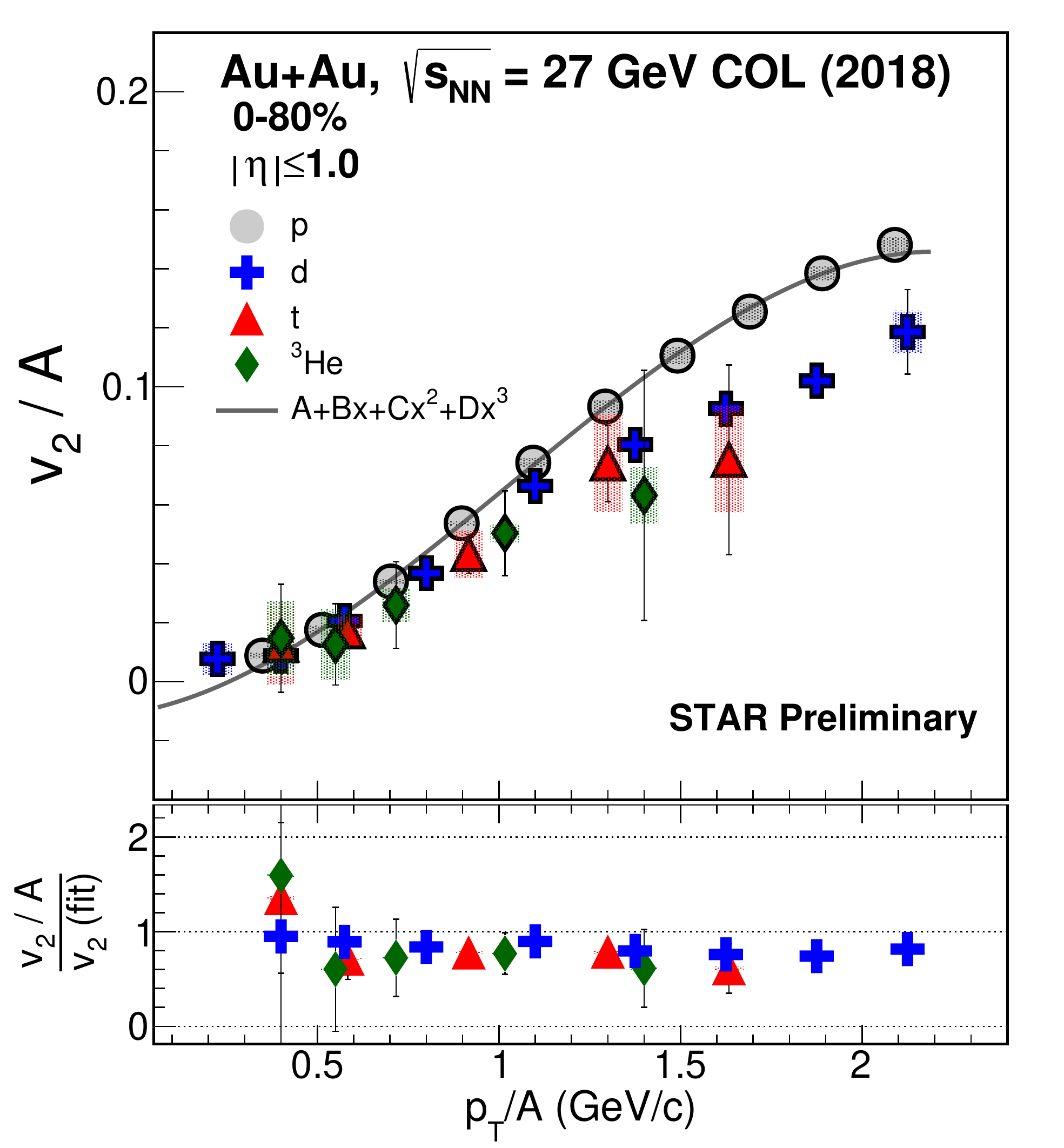}}
    \subfigure{\includegraphics[width=0.24\textwidth]{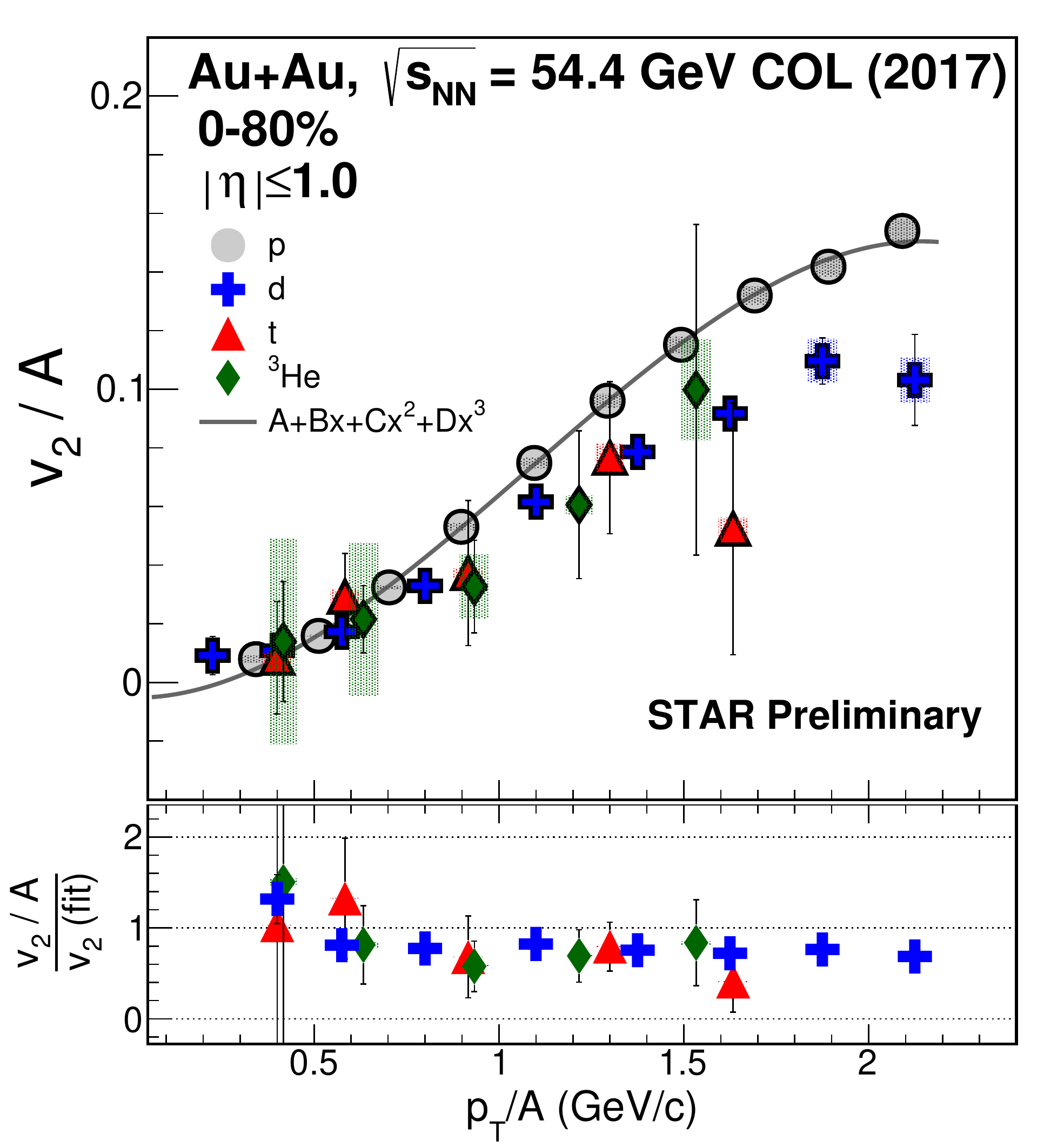}}
    \caption{Atomic mass number scaling $v_2/A$ of light nuclei as a function of $p_T/A$ in minimum bias Au+Au collisions at $\sqrt{s_{NN}}$ = 14.6, 19.6, 27, and 54.4 GeV. Proton $v_2$ has been fitted with a third-order polynomial. The bottom panel in each plot shows the ratio between the $v_2/A$ of light nuclei and the fit to proton $v_2$. Vertical lines and shaded area at each marker represent statistical and systematic uncertainties, respectively.}
    \label{fig:foobar}
\end{figure}

\subsection{Directed flow of hyper-nuclei}
Figure 4 shows $v_1$ as a function of rapidity ($y$) for $\Lambda$, $^{3}_{\Lambda}$H, and $^{4}_{\Lambda}$H in Au+Au collisions at $\sqrt{s_{NN}}$ = 3 GeV in 5-40\% centrality. Results of hyper-nuclei $v_1$ are compared to those of light nuclei $p$, $d$, $t$, $^3\text{He}$, and $^4\text{He}$. Linear fits (yellow-red lines) to $\Lambda$, $^{3}_{\Lambda}$H, and $^{4}_{\Lambda}$H $v_1$ are compared with the linear fits (gray lines) to light nuclei $v_1$.  These fit results have been extrapolated to the positive $y$ region. It is observed that the linear fit to the $\Lambda$ hyperon $v_1$ is consistent with that of proton, and the fit results of hyper-nuclei $v_1$ are also similar to those of the corresponding light nuclei with the similar mass number within statistical uncertainties. 
\begin{figure}[H]
    \centering
    \subfigure{\includegraphics[width=0.70\textwidth]{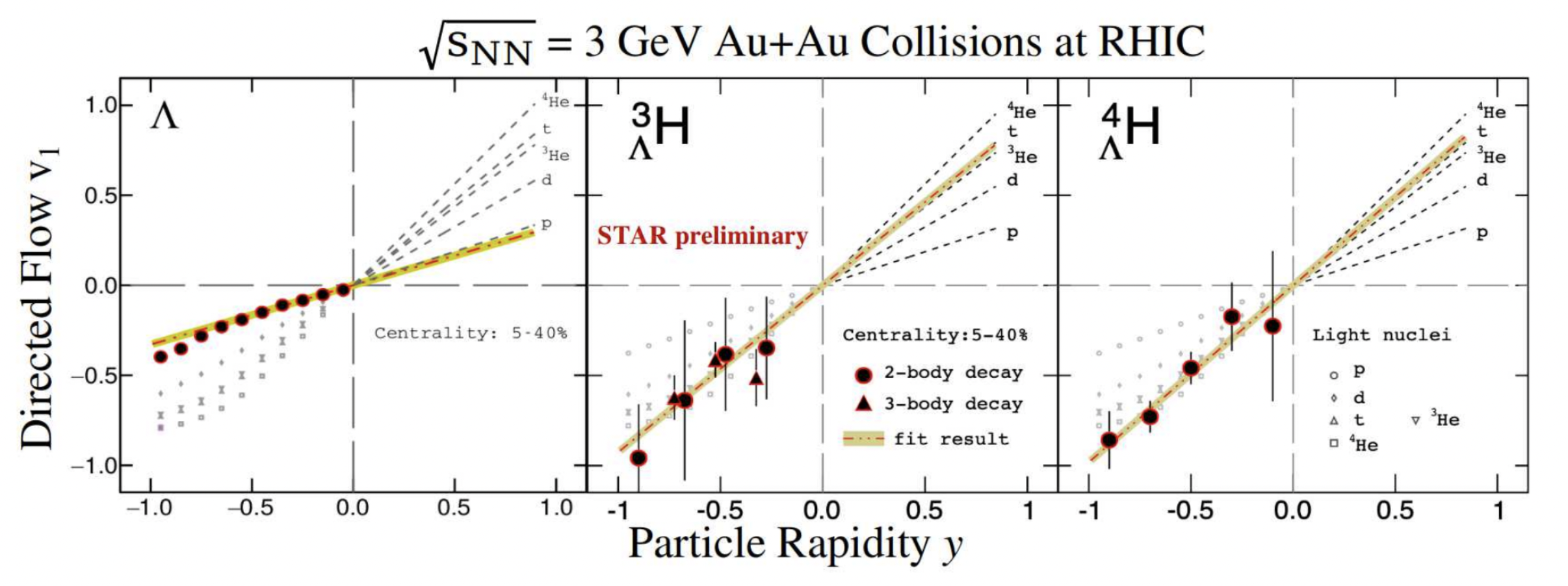}} 
    \caption{$v_1$ of $\Lambda$, $^{3}_{\Lambda}$H, and $^{4}_{\Lambda}$H as a function of rapidity from mid-central (5-40\%) Au+Au collisions at $\sqrt{s_{NN}}$ = 3 GeV. Vertical lines at each marker represent statistical uncertainty.}
    \label{fig:foobar}
\end{figure}

\begin{figure}

\centering
\sidecaption
\includegraphics[width=5cm,clip]{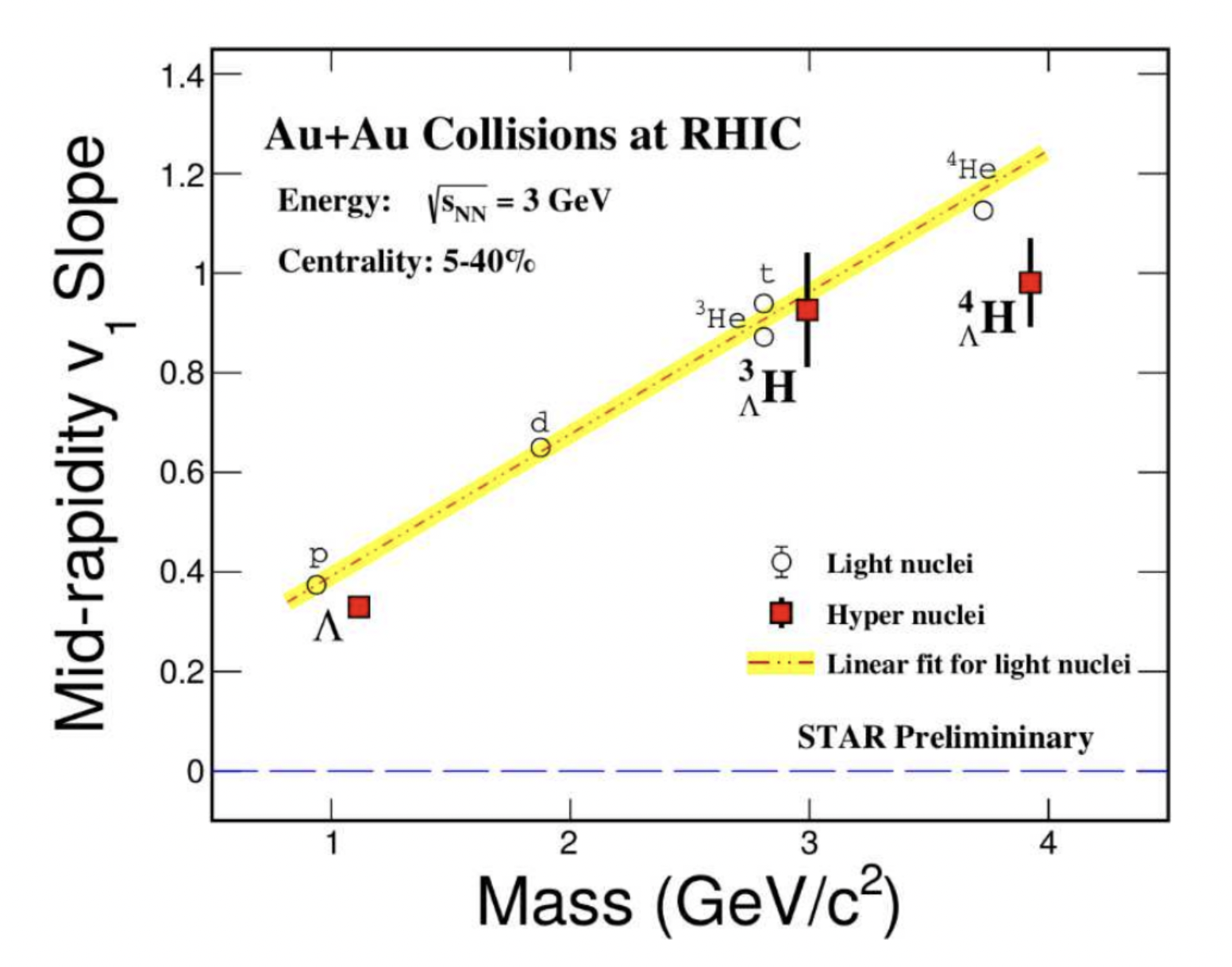}
\caption{Mass dependence of the mid-rapidity $v_1$ slope $dv_{1}/dy|_{y=0}$ for $\Lambda$, $^3_\Lambda$H, and $^4_\Lambda$H, from mid-central (5-40\%) Au+Au collisions at $\sqrt{s_{NN}}$ = 3 GeV. Vertical lines at each marker represent statistical uncertainty.}
\label{fig-3}     
\end{figure}
Mid-rapidity hyper-nuclei slopes of $v_1$ were extracted from the linear fits as shown in Fig. 4. Mass dependence of these $v_1$ slopes ($dv_1/dy$) is shown in Fig. 5. The $dv_1/dy$ of light nuclei $p$, $d$, $t$, $^3$He, and $^4$He from the same dataset and centrality is also shown as open circles. Linear fit to the light nuclei $dv_1/dy$ is shown as yellow-red line. It is observed that the $v_1$ slopes of hyper-nuclei are consistent with those of light nuclei with similar mass within statistical uncertainties. This mass dependence of the $v_1$ slope might suggest that coalescence could be the dominant mechanism of hyper-nuclei production in heavy-ion collisions.

\section{Summary}
In summary, we have studied the $v_2$ of $d$, $t$, and $^3\text{He}$ in Au+Au collisions at $\sqrt{s_{NN}}$ = 14.6, 19.6, 27, and 54.4 GeV. A monotonic rise of light nuclei $v_2$ with $p_T$ is observed. The $v_2$ of $d$ is observed to show a strong centrality dependence being higher for more peripheral collisions compared to more central collisions. Light nuclei $v_2$ is also found to follow the mass number scaling within 20-30\%. In addition, we have also reported the $v_1$ of $^{3}_{\Lambda}$H and $^{4}_{\Lambda}$H from Au+Au collisions at $\sqrt{s_{NN}}$ = 3 GeV. The rapidity dependence of hyper-nuclei $v_1$ and $dv_1/dy$ is compared to those of the light nuclei. It is observed that within statistical uncertainties, the results of hyper-nuclei $v_1$ and $dv_1/dy$ are quite close to that of corresponding light nuclei with similar mass. These observations may suggest that coalescence of nucleons and hyperons is the dominant mechanism of hyper-nuclei production in such collisions.


\begin{thebibliography}{}

\bibitem{RefCoal1}
S. T. Butler and C. A. Pearson, Phys. Rev. Lett., \textbf{129}, 836 (1963).
\bibitem{RefCoal2}
A. Schwarzschild and A. Zupancic, Phys. Rev. Lett., \textbf{129}, 854–862 (1963).
\bibitem{RefCoal3}
 H. H. Gutbrod et al., Phys. Rev. Lett., \textbf{37}, 667 (1976).
\bibitem{RefCoal4}
H. Sato and K. Yazaki, Phys. Lett. B, \textbf{98}, 153–157 (1981).
\bibitem{RefCoal5}
E. A. Remler, Annals of Physics, \textbf{136}, 293–316 (1981).
\bibitem{NS1}
D. Gerstung, N. Kaiser, and W. Weise, Eur. Phys. J., \textbf{A56}, 175 (2020).
\bibitem{NS2}
D. Lonardoni et al., Phys. Rev. Lett., \textbf{114}, 092301 (2015).

\bibitem{EPD}
J. Adams et al. (STAR Collaboration), Nucl. Instrum. Meth. A \textbf{968}, 163970 (2020).

\bibitem{MSN1}
T. Z. Yan, Y. G. Ma, X. Z. Cai et al., Phys. Lett. B, \textbf{638}, 50–54 (2006).
\bibitem{MSN2}
Y. Oh and C. M. Ko, Phys. Rev. C, \textbf{76}, 054910 (2007).

\end{thebibliography}
\end{document}